\documentclass[11pt]{article}

\usepackage{thumbpdf,lmodern}
\usepackage{makecell}
\usepackage{tabularx}
\usepackage{graphicx}
\usepackage{multirow}
\usepackage[round]{natbib}
\usepackage{amsfonts}
\usepackage{framed}
\usepackage{hyperref}
\usepackage{authblk}
\usepackage{tablefootnote}

\begin{document}
\sloppy

\title{\textbf{BayesPPD}: An \textbf{R} Package for Bayesian Sample Size Determination Using the Power and Normalized Power Prior for Generalized Linear Models}

\author[1]{Yueqi Shen\thanks{ys137@live.unc.edu}}
\author[1]{Matthew A. Psioda}
\author[1]{Joseph G. Ibrahim}
\affil[1]{Department of Biostatistics, University of North Carolina at Chapel Hill}

\maketitle

\begin{abstract}
The \textbf{R} package \textbf{BayesPPD} (Bayesian Power Prior Design) supports Bayesian power and type I error calculation and model fitting after incorporating historical data with the power prior and the normalized power prior for generalized linear models (GLM). The package accommodates summary level data or subject level data with covariate information. It supports use of multiple historical datasets as well as design without historical data. Supported distributions for responses include normal, binary (Bernoulli/binomial), Poisson and exponential. The power parameter $a_0$ can be fixed or modeled as random using a normalized power prior for each of these distributions. In addition, the package supports the use of arbitrary sampling priors for computing Bayesian power and type I error rates, and has specific features for GLMs that semi-automatically generate sampling priors from historical data. Since sample size determination (SSD) for GLMs is computationally intensive, an approximation method based on asymptotic theory has been implemented to support applications using the power prior. In addition to describing the statistical methodology and functions implemented in the package to enable SSD, we also demonstrate the use of \textbf{BayesPPD} in two comprehensive case studies. 
\end{abstract}

\section{Introduction: BayesPPD} 
\label{sec:intro}

There has been increasing interest over the past few decades in incorporating historical data in clinical trials, particularly on controls \citep{Pocock_1976, Neuenschwander_2010, Viele_2014}. Use of historical data can increase effective sample size, potentially leading to more accurate point estimates and increased power \citep{Neuenschwander_2010, Viele_2014}. Bayesian methods provide a natural mechanism for information borrowing through the use of informative priors. Some popular informative priors for Bayesian clinical trial design include the power prior \citep{chen_2000}, the normalized power prior \citep{duan_2006}, the commensurate power prior \citep{Hobbs_2011}, and the robust meta-analytic-predictive prior \citep{Schmidli_2014}.

Some advantages of the power prior include its easy construction, its natural way of incorporating historical data, its intuitive interpretation, and its desirable theoretical properties \citep{ibrahim_2015}. For example,  \cite{ibrahim_2003} show that the power prior is an optimal class of informative priors in the sense that it minimizes a convex sum of the Kullback–Leibler (KL) divergences between two posterior densities, in which one density is based on no incorporation of historical data, and the other density is based on pooling the historical and current data. \cite{duan_2006} propose a modification of the power prior, the normalized power prior, which adds a normalizing constant component when the power parameter is modeled as random. The normalizing constant poses computational challenges in the presence of covariates, because it is analytically intractable except in the case of the normal linear model. We address this challenge by utilizing the PWK estimator \citep{pwk_2018} to approximate the normalizing constant for use with generalized linear models. We also develop a novel way of incorporating the approximation of the normalizing constant into the Markov chain Monte Carlo (MCMC) algorithm.

There is a growing literature on Bayesian sample size determination, including the works of \cite{Rahme_1998}, \cite{Simon_1999}, \cite{Gelfand_2002}, \cite{DeSantis_2007}, \cite{MLan_2006} and \cite{MLan_2008}. We consider the simulation-based method developed in \cite{Chen_2011} and \cite{Psioda_Ibrahim_2019}, which extends the the fitting and sampling priors of \cite{Gelfand_2002} with a focus on controlling the type I error rate and calculating power. In addition, our package supports the use of arbitrary sampling priors for computing Bayesian power and type I error rates, and has specific features for GLMs that semi-automatically generate sampling priors from historical data.

The \textbf{R} \citep{R_2017} package \textbf{BayesPPD} (Bayesian Power Prior Design) \citep{BayesPPD} supports Bayesian clinical trial design after incorporating historical data with the power prior and the normalized power prior. \textbf{BayesPPD} has two categories of functions: functions for model fitting and functions for Bayesian power and type I error rate estimation. The package accommodates summary level data or subject level data with covariate information for normal, binary (Bernoulli/binomial), Poisson and exponential models. It supports use of multiple historical datasets and design without historical data. 

Several Bayesian clinical trial design packages are available on the Comprehensive R Archive Network (CRAN), such as \textbf{BAEssd}, \textbf{BDP2}, \textbf{ph2bayes} and \textbf{gsbDesign} \citep{BAEssd, BDP2, ph2bayes, gsbDesign}. However, these packages do not accommodate the incorporation of historical data and are limited to normal and binary endpoints. The \textbf{RBesT} package \citep{RBesT} accounts for historical data using the meta-analytic-predictive prior. Commercial software for clinical trial design such as \textbf{FACTS}, \textbf{East} and \textbf{ADDPLAN} \citep{FACTS, East, ADDPLAN} do not implement the power prior, to our knowledge.  The \textbf{BayesCTDesign} \citep{BayesCTDesign} package supports two-arm randomized Bayesian trial design using historical control data with the power prior, but it does not allow covariates, nor does it allow the power parameter to be treated as random.  The \textbf{NPP} \citep{NPP} package implements the normalized power prior for two group cases for Bernoulli, normal, multinomial and Poisson models, as well as for the normal linear model. It does not support generalized linear models, nor does it include functions for sample size determination. The \textbf{bayesDP} \citep{bayesDP} package implements the power prior where the power parameter is determined by a discounting function estimated based on a measure of prior-data conflict. Thus, this approach is not fully Bayesian, and the package must be used in conjunction with the package \textbf{bayesCT} \citep{bayesCT} for trial design. While \textbf{bayesDP} supports two-arm trials for binomial, normal and survival models as well as linear and logistic regression models, \textbf{BayesPPD} allows covariates for Bernoulli/binomial, normal, Poisson and exponential models with several choices of link functions. The \textbf{BayesPPD} package is a comprehensive resource that supports Bayesian analysis and design using the power prior and normalized power prior.

Another advantage of \textbf{BayesPPD} is its computational speed.  \textbf{BayesPPD} implements MCMC algorithms with \textbf{Rcpp} \citep{Rcpp} without recourse to asymptotics. For most sample sizes, functions for analysis take only a few seconds to run. Functions for design for two group cases run in seconds for fixed $a_0$, and generally run in less an hour for random $a_0$, depending on the desired level of precision (e.g., number of simulated datasets). In the presence of covariates, functions for design are more computation-intensive; an approximation method based on asymptotic theory has been implemented to help users obtain a rough estimate of the desired sample size before fine-tuning using the MCMC-based method. 

This article is organized as follows. Section \ref{sec:tf} describes the methods implemented by the package. Section \ref{sec:use} provides details on how to use \textbf{BayesPPD} for different data scenarios and model needs. Section \ref{sec:ex} presents two case studies with example code, one with covariates and one without. The article is concluded with a brief discussion.

\section[Theoretical Framework]{Theoretical framework} \label{sec:tf}

\subsection{Basic formulation of the power prior}

Let $D$ denote data from the current study and $D_0$ denote data from a historical study. Let $\theta$ denote model parameters and $L(\theta|D)$ denote a general likelihood function associated with a given outcome model, such as a linear model, generalized linear model (GLM), survival model, or random effects model. Following \cite{chen_2000}, the power prior is formulated as $$\pi(\theta|D_0, a_0) \propto L(\theta|D_0)^{a_0}\pi_0(\theta).$$ where $0 \le a_0 \le 1$ is a  discounting parameter for the historical data likelihood, and $\pi_0(\theta)$ is the initial prior for $\theta$. The parameter $a_0$ allows researchers to control the influence of the historical data on the posterior distribution. When $a_0=0$, historical information is discarded and the power prior becomes equivalent to the initial prior $\pi_0(\theta)$. When $a_0=1$, the power prior corresponds to the posterior distribution of $\theta$ given the historical data and the initial prior. When $a_0$ is treated as fixed, sensitivity analysis can be performed to determine an appropriate $a_0$ value. When $a_0$ is treated as random, priors such as the beta distribution can be specified. The choice of $a_0$ is discussed in, for example,  \cite{ibrahim_2015} and \cite{psioda_2018}.   

The power prior can easily accommodate multiple historical datasets. Suppose there are $K$ historical datasets denoted by $D_{0k}$ for $k=1,\cdots, K$ and let $D_0=(D_{01}, \cdots, D_{0K})$. The power prior becomes $$\pi(\theta|D_0, a_0) \propto \prod_{k=1}^K L(\theta|\\D_{0k})^{a_{0k}}\pi_0(\theta)$$ where $a_0 = (a_{01},\cdots,a_{0K})'$ and $0\le a_{0k} \le 1$ for $k=1,\cdots,K$.

\subsection{The normalized power prior}

Modeling $a_0$ as random allows one to represent uncertainty in how much the historical data should be discounted. The simplest power prior that allows this is the \emph{joint power prior} \citep{chen_2000} which is given by $$\pi(\theta, a_0|D_0) \propto L(\theta|D_0)^{a_0}\pi_0(\theta)\pi_0(a_0).$$ \cite{neuens_2009} point out that this formulation is not ideal because the normalizing constant $$c(a_0)=\int L(\theta|D_0)^{a_0}\pi_0(\theta) d\theta$$ for $L(\theta|D_0)^{a_0}\pi_0(\theta)$ is not incorporated and thus $\pi_0(a_0)$ is not actually the marginal prior for $a_0$. In fact, \cite{duan_2006} point out that this formulation of the power prior does not obey the likelihood principle. \cite{duan_2006} proposed a modification of the power prior, the \emph{normalized power prior}, which is given by $$\pi(\theta, a_0|D_0) = \pi(\theta|D_0, a_0)\pi(a_0) = \frac{L(\theta|D_0)^{a_0}\pi_0(\theta)}{c(a_0)}\pi_0(a_0),$$ where $\pi_0(a_0)$ is the initial prior for $a_0$. The normalized power prior specifies a conditional prior for $\theta$ given $a_0$ and a marginal prior for $a_0$. The normalizing constant $$c(a_0)=\int L(\theta|D_0)^{a_0}\pi_0(\theta) d\theta$$ is often analytically intractable and requires Monte Carlo methods for estimation. When $a_0$ is modeled as random, the normalized power prior is implemented in \textbf{BayesPPD} using a beta initial prior on $a_0$, for which the user must specify values of the two shape parameters that define the beta density. The package supports the inclusion of multiple historical datasets when $a_0$ is modeled as random. 

\subsection{The power prior for generalized linear models}
The power prior can easily accommodate covariates. Let $y_i$ denote the response variable and $x_i$ denote a $p$-dimensional vector of covariates for subject $i=1, \cdots, n$. Denote $\tilde{\beta}=(\beta_0, \beta)$, where $\beta_0$ is the intercept and $\beta = (\beta_1, \cdots, \beta_p)'$ is a $p$-dimensional vector of regression coefficients. We assume the GLM of $y_i|x_i$ is given by $$f(y_i|x_i, \tilde{\beta}, \tau) = \exp\{\alpha_i^{-1}(\tau)(y_ig(\beta_0+x_i'\beta)-\psi(g(\beta_0+x_i'\beta))) + \phi(y_i, \tau)\}, $$ where $\tau$ is a scale parameter and $g$ is a monotone differentiable link function. In particular, \textbf{BayesPPD} allows the distribution of $y_i|x_i$ to be normal, Bernoulli, binomial, Poisson or exponential. Note that for Bernoulli, binomial, Poisson and exponential regression models, $\tau$ is equal to $1$. 

Let $D_{0k} = \{(y_{0ki}, x_{0ki}), i=1,\cdots, n_{0k}\}$ denote the $k$-th historical dataset, where $y_{0ki}$ is the response variable for historical control $i$ and $x_{0ki}$ is the $p$-dimensional vector of covariates for historical control $i$. The GLM for $y_{0ki}|x_{0ki}$ is $$f(y_{0ki}|x_{0ki}, \tilde{\beta}, \tau_{0k}) = \exp\{\alpha_{0i}^{-1}(\tau_{0k})(y_{0ki}g(\beta_0+x'_{0ki}\beta)-\psi(g(\beta_0+x'_{0ki}\beta))) + \phi(y_{0ki}, \tau_{0k})\},$$ where $\tau_{0k}$ is the scale parameter for the $k$-th historical dataset. Note that the precision parameter is assumed to be unshared. The historical data likelihood for $K$ historical datasets is $L(\tilde{\beta}, \tau_{01}, \cdots, \tau_{0K}|D_0) \propto \prod_{k=1}^{K}\prod_{i=1}^{n_{0k}}f(y_{0ki}|x_{0ki}, \tilde{\beta}, \tau_{0k})$. The power prior for GLMs with fixed $a_0 = (a_{01},\cdots,a_{0K})'$ is $$\pi(\tilde{\beta}, \tau_{01}, \cdots, \tau_{0K}|D_0, a_0) \propto \prod_{k=1}^{K}\left\{L(\tilde{\beta}, \tau_{0k}|D_{0k})^{a_{0k}}\pi_0(\tau_{0k})\right\}\pi_0(\tilde{\beta}).$$ When $a_0$ is modeled as random, we assume $\tau_{01},\cdots,\tau_{0K}=\tau$ for computational simplicity. The normalized power prior for GLMs with a random $a_0$ vector is given by $$\pi(\tilde{\beta}, \tau, a_0|D_0) = \frac{\prod_{k=1}^{K}L(\tilde{\beta}, \tau|D_{0k})^{a_{0k}}\pi_0(\tilde{\beta})\pi_0(\tau)}{\int_0^\infty\int_{\mathbb{R}^p} \prod_{k=1}^{K}L(\tilde{\beta}, \tau|D_{0k})^{a_{0k}}\pi_0(\tilde{\beta})\pi_0(\tau)d\tilde{\beta} d\tau} \pi_0(a_0).$$

\subsection{Estimating the normalizing constant for GLMs}

The normalizing constant $c(a_0)$ in the normalized power prior for GLMs is analytically intractable except for normal linear regression models. For other types of regression models, we approximate the normalizing constant with the partition weighted kernel (PWK) estimator proposed by \cite{pwk_2018}. The PWK estimator requires MCMC samples from the posterior distribution (based on a discounted historical data likelihood with fixed $a_0$ value), which we obtain using the slice sampler \citep{slice}, and the known kernel function for computing the normalizing constant. The authors first impose a working parameter space, defined as the space where the kernel value is bounded away from zero. As stated in \cite{pwk_2018}, the PWK estimator is constructed by first partitioning the working parameter space and then estimating the marginal likelihood by a weighted average of the kernel values evaluated at a MCMC sample for each partition, where the weights are assigned locally using a representative kernel value in each partitioned subset. The PWK estimator has been shown to have desirable properties, including being consistent and having finite variance \citep{pwk_2018}. 

The function \textit{normalizing.constant} in our package computes a vector of coefficients that defines a function $f(a_0)$ that approximates the normalizing constant for GLMs with random $a_0$. Suppose there are $K$ historical datasets. Basic usage of the \textit{normalizing.constant} function entails the following steps:

\begin{enumerate}
\item  The user inputs a grid of $M$ rows and $K$ columns of potential values for $a_0$. 
\item For each row of $a_0$ values in the grid, the function obtains $M$ samples for $\beta$ from the power prior associated with the current values of $a_0$ using the slice sampler. Note that $\tau$ is not applicable here because the models implemented using the PWK estimator do not have scale parameters. 
\item For each of the $M$ sets of posterior samples, the PWK algorithm \citep{pwk_2018} is used to estimate the log of the normalizing constant $d_1,\cdots, d_M$ for the normalized power prior. 
\item At this point, one has a dataset with outcomes $d_1,\cdots, d_M$ and predictors corresponding to the rows of the $a_0$ grid matrix. A polynomial regression is employed to estimate a function $d = f(a_0)$ based on these quantities. The degree of the polynomial regression is determined by the algorithm to ensure $R^2 > 0.99$.
\item The \textit{normalizing.constant} function returns the vector of coefficients from the polynomial regression model, which the user must input into the analysis or design function for GLMs with $a_0$ modeled as random (\textit{glm.random.a0} and \textit{power.glm.random.a0}).
\end{enumerate}

\subsection{Sample size determination}
\subsubsection{Hypotheses for two group models}

Following \cite{Chen_2011}, for two group models (i.e., treatment and control group with no covariates), denote the parameter for the treatment group by $\mu_t$ and the parameter for the control group by $\mu_c$. For example, for binomial models, $\mu_t$ and $\mu_c$ are the probability of having some outcome (e.g., tumor response) for the treatment and control group, respectively. Let $\tau_c$ denote the nuisance parameters for the control group in the model. For normal models, $\tau_c$ is a vector of precision parameters. For $K$ historical datasets $D_0 = (D_{01},\cdots, D_{0K})'$ with fixed $a_0$, we assume each historical dataset $D_{0k}$ has a precision parameter $\tau_{c0k}$. When $a_0$ is modeled as random, the historical and current datasets are assumed to have the same precision parameter, in which case $\tau_c$ reduces to a scalar. The precision parameter of the treatment group is denoted by $\tau_t$. 

We consider the following power prior for ($\mu_c$, $\tau_c$) given multiple historical datasets $D_0$ $$\pi(\mu_c, \tau_c|D_0,a_0) \propto \prod_{k=1}^K \left[L(\mu_c|D_{0k}, \tau_c)^{a_{0k}}\right]\pi_0(\mu_c)\pi_0(\tau_c),$$ where $a_0 = (a_{01},\cdots,a_{0K})'$, $0\le a_{0k} \le 1$ for $k=1,\cdots,K$, $L(\mu_c|D_{0k}, \tau_c)$ is the historical data likelihood, and $\pi_0(\mu_c)$ and $\pi_0(\tau_c)$ are the initial priors. To model $a_0$ as random, we consider the normalized power prior $$\pi(\mu_c, \tau_c, a_0|D_0) \propto \frac{\prod_{k=1}^K \left[L(\mu_c|D_{0k}, \tau_c)^{a_{0k}}\right]\pi_0(\mu_c)\pi_0(\tau_c)}{c(a_0)}\pi_0(a_0),$$ where $$c(a_0)=\int_0^\infty\int_{-\infty}^\infty \prod_{k=1}^K [L(\mu_c|D_{0k}, \tau_c)^{a_{0k}}]\pi_0(\mu_c)\pi_0(\tau_c)d\mu_cd\tau_c.$$

For models other than the exponential model, the power / type I error calculation algorithm assumes the null and alternative hypotheses are given by $$H_0: \mu_t - \mu_c \ge \delta$$ and $$H_1: \mu_t - \mu_c < \delta,$$ where $\delta$ is a prespecified constant. To test hypotheses of the opposite direction, i.e., $H_0: \mu_t - \mu_c \le \delta$ and $H_1: \mu_t - \mu_c > \delta$, one can recode the responses for the treatment and control groups. 

For positive continuous data assumed to follow exponential distribution, the hypotheses are given by
$$H_0: \mu_t/\mu_c \ge \delta$$ and $$H_1: \mu_t/\mu_c < \delta,$$ where $\mu_t$ and $\mu_c$ are the hazards for the treatment and the control group, respectively.

\subsubsection{Definition of Bayesian type I error rate and power}

Let $\Theta_0$ and $\Theta_1$ denote the parameter spaces corresponding to $H_0$ and $H_1$. Let $y^{(n)}$ denote the simulated current data associated with a sample size of $n$ and let $\theta=(\mu_t, \mu_c, \tau_c)$ denote the model parameters. Let $\pi^{(s)}(\theta)$ denote the sampling prior and let $\pi^{(f)}(\theta)$ denote the fitting prior. The sampling prior is used to generate the hypothetical data while the fitting prior is used to fit the model after the data is generated. Let $\pi_0^{(s)}(\theta)$ denote a sampling prior that only puts mass in the null region, i.e., $\theta \subset \Theta_0$. Let $\pi_1^{(s)}(\theta)$ denote a sampling prior that only puts mass in the alternative region, i.e., $\theta \subset \Theta_1$. To determine Bayesian sample size, we estimate the quantity $$\beta_{sj}^{(n)}=E_s[I\{P(\mu_t-\mu_c<\delta|y^{(n)}, \pi^{(f)})\ge \gamma\}]$$ where $j=0$ or $1$, corresponding to the expectation taken with respect to $\pi_0^{(s)}(\theta)$ or $\pi_1^{(s)}(\theta)$. The constant $\gamma > 0$ is a prespecified posterior probability threshold for rejecting the null hypothesis (e.g., $0.975$). The probability is computed with respect to the posterior distribution given the simulated data $y^{(n)}$ and the fitting prior $\pi^{(f)}(\theta)$, and the expectation is taken with respect to the marginal distribution of $y^{(n)}$ defined based on the sampling prior $\pi^{(s)}(\theta)$. Then $\beta_{s0}^{(n)}$ corresponding to $\pi^{(s)}(\theta)=\pi_0^{(s)}(\theta)$ is the Bayesian type I error rate, while $\beta_{s1}^{(n)}$ corresponding to $\pi^{(s)}(\theta)=\pi_1^{(s)}(\theta)$ is the Bayesian power. Note that Bayesian type I error rate and power can be equivalently defined as weighted averages of the quantities based on fixed values of $\theta$ with weights determined by the sampling priors \citep{psioda_2018}. For given $\alpha_0 > 0$ and $\alpha_1 > 0$, we can compute $n_{\alpha_0} = \min\{n: \beta_{s0}^{(n)} \le \alpha_0\}$ and $n_{\alpha_1} = \min\{n: \beta_{s1}^{(n)} \ge 1-\alpha_1\}$. Then, the sample size is taken to be max$\{n_{\alpha_0}, n_{\alpha_1}\}$. Common choices of $\alpha_0$ and $\alpha_1$ include $\alpha_0=0.05$ and $\alpha_1=0.2$. These choices guarantee that the Bayesian type I error rate is at most $0.05$ and the Bayesian power is at least $0.8$.

\subsubsection{Estimation of Bayesian type I error rate and power}

In this section, we discuss the simulation-based procedure used to estimate the Bayesian type I error rate and power. Let $N$ denote the number of simulated trials. To compute $\beta_{sj}^{(n)}$, the following algorithm is used for each simulated trial $b$:
\begin{itemize}
\item{Step 1: }{Generate $\theta^{(b)} \sim \pi_j^{(s)}(\theta)$.}
\item{Step 2: }{Generate $y^{(b)} \sim f(y^{(b)}|\theta^{(b)})$.}
\item{Step 3: }{Estimate the posterior distribution $\pi(\theta|y^{(b)}, D_0, a_0)$ and the posterior probability $P(\mu_t - \mu_c < \delta|y^{(b)}, \pi^{(f)},D_0, a_0)$.}
\item{Step 4: }{Compute the indicator $r^{(b)}=I\{P(\mu_t - \mu_c < \delta|y^{(b)}, \pi^{(f)}, D_0, a_0) \ge \gamma$\}.}
\end{itemize}

Then the estimate of $\beta_{sj}^{(n)}$ is $\frac{1}{N}\sum_{b=1}^N r^{(b)}$.

\subsubsection{Specification for regression models}
For regression models, we assume the first column of the covariate matrix is the treatment indicator,
and the corresponding parameter is $\beta_1$, which, for example, corresponds to a difference in means for the linear regression model and a log hazard ratio for the exponential regression model.
The hypotheses are given by
$$H_0: \beta_1 \ge \delta$$ and $$H_1: \beta_1 < \delta.$$
The definition of $\beta_{sj}^{(n)}$ and the algorithm change accordingly.

\subsection{Prior distributions}

\subsubsection{Two group cases}

For two group models, continuous responses of the control group are assumed to follow $N(\mu_c, \tau_c^{-1})$. Each historical dataset $D_{0k}$ is assumed to have a different precision parameter $\tau_{c0k}$. The initial prior for the $\mu_c$ is the uniform improper prior. The initial prior for $\tau_c$ is the Jeffery’s prior, $\tau_c^{-1}$, and the initial prior for $\tau_{c0k}$ is $\tau_{c0k}^{-1}$. Posterior samples of $\mu_c$, $\tau_c$ and $\tau_{c0k}$'s (if historical data is given) are obtained through Gibbs sampling. When $a_0$ is modeled as random, the historical datasets are assumed to have the same precision parameter $\tau_c$ as the current dataset for computational simplicity. The initial prior for $\tau_c$ is the Jeffery’s prior, $\tau_c^{-1}$. Posterior samples of $a_0$ are obtained through slice sampling. 

For binary, count or positive continuous data, a single response from the control group is assumed to follow Bernoulli($\mu_c$), Poisson($\mu_c$) or exponential(rate=$\mu_c$), respectively. A beta initial prior is used for $\mu_c$ for Bernoulli data, and a gamma prior is used for Poisson and exponential data. The user can specify the hyperparameters. When $a_0$ is modeled as random, posterior samples of $a_0$ are obtained through slice sampling. The conditional posterior distributions of $\mu_c$ given $a_0$ have closed form solutions.

When computing the power or the type I error rate, treatment group data are simulated and posterior samples of $\mu_t$ (and $\tau_t$ for normal data) are obtained using basic Bayesian models. The priors used for $\mu_t$ are the same as the initial priors used for $\mu_c$. For normal data, the prior for $\tau_t$ is the Jeffery’s prior, $\tau_t^{-1}$.

\subsubsection{GLM cases}
For GLMs, a continuous response $y_i$ is assumed to follow $N(\beta_0+x_i'\beta, \tau^{-1})$. Each historical dataset $D_{0k}$ is assumed to have a different precision parameter $\tau_k$. The initial prior for $\tau$ is the Jeffery’s prior, $\tau^{-1}$, and the initial prior for $\tau_k$ is $\tau_k^{-1}$.  Posterior samples of $\beta_0$ and $\beta$ are obtained through Gibbs sampling. For all other types of data, a link function must be applied. Posterior samples of $\beta_0$ and $\beta$ are obtained through slice sampling. The initial prior for $\beta_0$ and $\beta$ is the uniform improper prior. When $a_0$ is modeled as random, the historical datasets are assumed to have the same precision parameter $\tau$ as the current dataset. The initial prior for $\tau$ is the Jeffery’s prior, $\tau^{-1}$. Posterior samples of $a_0$ are obtained through slice sampling. The normalizing constant of the normalized
power prior is estimated using the PWK estimator (see section \ref{sec:useGLM}).

\section{Using BayesPPD}
\label{sec:use}

\subsection{Package overview}
The \textbf{BayesPPD} package accommodates summary level data or subject level data with covariate information. It supports SSD for design applications with multiple historical datasets as well as with no historical data. Functions with names containing \textit{two.grp} assume that the input data are sufficient statistics (e.g., sample mean) for independent and identically distributed treatment and control group data. Simulated control group data are analyzed using the power or normalized power prior and posterior samples of $\mu_c$ are returned. Functions with names containing \textit{glm} assume that the historical control data include a covariate matrix $X_0$ and the current data include the same set of covariates with an additional column (the first column) of treatment indicator. Simulated data are analyzed using the power or normalized power prior and posterior samples of the regression coefficients are returned. For each of two cases, the power parameter $a_0$ can be fixed or modeled as random, resulting in four model fitting functions, \textit{two.grp.fixed.a0}, \textit{two.grp,random.a0}, \textit{glm.fixed.a0} and \textit{glm.random.a0}. For each of the four model fitting functions, a companion function prefixed with \textit{power} calculates power or type I error rate, given historical data and current data sample size. Supported distributions of responses include normal, binary (Bernoulli/binomial), Poisson and exponential. Since functions for sample size determination for GLMs are computationally intensive, an approximation method based on asymptotic theory has been implemented for the model with fixed $a_0$. 

Table~\ref{table:methods} shows the sampling methods used for each model and data distribution. Gibbs sampling is used for normally distributed data. Slice sampling \citep{slice} is used for all other data distributions, and for obtaining posterior samples of $a_0$ when $a_0$ is considered random. For two group models with fixed $a_0$, numerical integration is performed using the \textbf{RcppNumerical} package \citep{rcppnum}. For GLMs with random $a_0$, the PWK estimator \citep{pwk_2018} is used to estimate the normalizing constant. 

\begin{table}[t!]
\centering
\setlength{\extrarowheight}{5pt}
\begin{tabular}{ccccc}
\hline
 & \thead{Two groups,\\ fixed $a_0$} & \thead{Two groups,\\ random $a_0$} & \thead{GLM,\\ fixed $a_0$*} & \thead{GLM,\\ random $a_0$}\\[8pt]
\hline
\thead{Bernoulli/\\Binomial} & \thead{Numerical\\ integration} & Slice & Slice & Slice \& PWK \\ 
Normal & Gibbs & Gibbs \& Slice  & Gibbs & Gibbs \& Slice\\
Poisson & \thead{Numerical\\ integration} & Slice & Slice &Slice \& PWK \\
Exponential & \thead{Numerical\\ integration} & Slice  & Slice & Slice \& PWK\\
\hline
\end{tabular}
\caption{Estimation method used for each model and data type.\\ $^*$Approximation method is available for sample size determination for fast implementation.}
\label{table:methods}
\end{table}

\subsection{Two group cases}

If one has current and/or historical control data for an application with no covariates and would like to obtain posterior samples of $\mu_c$ (and $\tau_c$ for normal data), one uses the function \textit{two.grp.fixed.a0} or \textit{two.grp.random.a0}. The user must specify the \textit{data.type} (\textit{"Normal"}, \textit{"Bernoulli"}, \textit{"Poisson"} or \textit{"Exponential"}), the sum of responses \textit{y.c}, the sample size \textit{n.c} and the sample variance \textit{v.c} (for normal data only) of the current control data. The optional \textit{historical} argument is a matrix where the columns contain the sufficient statistics and each row represents a historical dataset. For \textit{two.grp.fixed.a0}, \textit{historical} must contain a column of $a_0$ values, one $a_0$ value for each historical dataset. For non-normal data, the user can specify \textit{prior.mu.c.shape1} and \textit{prior.mu.c.shape2}, the hyperparameters of the initial prior for $\mu_c$. 

When $a_0 = (a_{01},\cdots,a_{0K})'$ is modeled as random, a beta prior is specified for $a_0$ with hyperparameters \textit{prior.a0.shape1} and \textit{prior.a0.shape2}. Posterior samples of $a_0$ are obtained through slice sampling. The optional tuning parameters for the slice sampler include \textit{lower.limits} and \textit{upper.limits} which control the upper and lower limits of the parameters being sampled, as well as \textit{slice.widths} which controls the width of each slice. The length of \textit{lower.limits}, \textit{upper.limits} and \textit{slice.widths} should be at least equal to the number of parameters, i.e., the dimension of $a_0$. Their default values are $0$, $1$ and $0.1$, respectively, for each $a_{0k}$.

For sample size determination, \textit{power.two.grp.fixed.a0} and \textit{power.two.grp.random.a0} compute the power or the type I error rate given the sample sizes of the treatment and control groups for the new study and other inputs. If a sampling prior with support in the null space is used, the value returned is a Bayesian type I error rate. If a sampling prior with support in the alternative space is used, the value returned is a Bayesian power. The arguments \textit{samp.prior.mu.t} and \textit{samp.prior.mu.c} contain vectors of samples for $\mu_t$ and $\mu_c$, which are discrete approximations of the sampling priors. For normal data, arguments \textit{samp.prior.var.t} and \textit{samp.prior.var.c}, which contain samples for $\tau_t^{-1}$ and $\tau_c^{-1}$, must also be provided. Section \ref{sec:sp} details the choice of sampling priors. The argument \textit{delta} specifies the constant that defines the boundary of the null hypothesis. The default value is zero. The argument \textit{gamma} specifies the posterior probability threshold for rejecting the null hypothesis. The default value is 0.95.

\subsection{GLM cases}\label{sec:useGLM}

If one has current and/or historical control data for an application with covariates and would like to obtain posterior samples of $\beta$ (and $\tau$ for normal data), one uses the function \textit{glm.fixed.a0} or \textit{glm.random.a0}. It is recommended that the covariates be transformed or standardized so that the estimation of $\beta$ will be stable. The user must specify the \textit{data.type}, the \textit{data.link} (except for normal data), the vector of responses \textit{y} and the matrix of covariates \textit{x} where the first column should be the treatment indicator. Supported link functions include logit, probit, log, identity-positive, identity-probability
and complementary log-log. If the data is binary and all covariates are discrete, the user can collapse the  Bernoulli data into a binomial structure, which may result in a much faster slice sampler. In this case, the user needs to provide \textit{n}, a vector of integers specifying the number of subjects who have a particular value of the covariate vector. The optional \textit{historical} argument is a list of lists where each list contains information about a historical dataset with named elements \textit{y0}, \textit{x0} and \textit{a0} (only for \textit{glm.fixed.a0}). The historical covariate matrix \textit{x0} should not have the treatment indicator since we assume only historical control data is available. Apart from missing the treatment indicator, \textit{x0} should have the same set of covariates in the same order as \textit{x}. For non-normal data, slice sampling is used to obtain posterior samples of $\beta$, and the user can specify the \textit{lower.limits}, \textit{upper.limits} and \textit{slice.widths} of the sampler. The length of \textit{lower.limits}, \textit{upper.limits} and \textit{slice.widths} should be at least equal to the number of parameters, i.e., the dimension of $\beta$. A matrix of posterior samples of $\beta$ is returned, where the first column contains posterior samples of the intercept and the second column contains posterior samples of $\beta_1$, the parameter for the treatment indicator.

When $a_0$ is modeled as random for non-normal data, the user must first use the function \textit{normalizing.constant} to obtain the value of \textit{a0.coefficients}, a vector of coefficients for $a_0$ necessary for estimating the normalizing constant for the normalized power prior. For the \textit{grid} argument of \textit{normalizing.constant}, the user inputs a grid of $M$ rows and $K$ columns of potential values for $a_0$ for $K$ historical datasets. For example, one can choose the vector \textit{v = c(0.1, 0.25, 0.5, 0.75, 1)} and use \verb!expand.grid(a0_1=v, a0_2=v, a0_3=v)! when $K=3$ to get a grid with $M = 5^3 = 125$ rows and three columns. If there are more than three historical datasets, the dimension of $v$ can be reduced to limit the size of the grid. A large grid will increase runtime. If some of the coefficients are not estimable in the polynomial regression, the algorithm will product the error message, "some coefficients not defined because of singularities." To resolve the issue, the user can try increasing or decreasing the number of rows in the grid. Other possible causes include insufficient sample size of the historical data, insufficient number of iterations for the slice sampler, and near-zero grid values.

When $a_0$ is modeled as random, slice sampling is used for $a_0$ only for normal data, and the length of \textit{lower.limits}, \textit{upper.limits} and \textit{slice.widths} should be equal to the dimension of $a_0$. For all other data types, slice sampling is used for $\beta$ and $a_0$, and the length of those vectors should be equal to the dimension of $\beta$ plus the dimension of $a_0$. 

For sample size determination, \textit{power.glm.fixed.a0} and \textit{power.glm.random.a0} compute the power or the type I error given the total sample size (\textit{data.size}) for the new study and other inputs. If historical datasets are provided, the algorithm samples with replacement from the historical covariates to construct the simulated datasets. Otherwise, the algorithm samples with replacement from \textit{x.samples}. One of the arguments \textit{historical} and \textit{x.samples} must be provided. The argument \textit{samp.prior.beta} contains a matrix of samples for $\beta$, which is a discrete approximation of the sampling prior. For normal data, the argument \textit{samp.prior.var} containing samples for $\tau^{-1}$ must also be provided. The average posterior means of the parameters are also returned.

\subsection{Sampling priors}
\label{sec:sp}

Our implementation in \textbf{BayesPPD} does not assume any particular distribution for the sampling priors. The user specifies discrete approximations of the sampling priors by providing a vector or a matrix of sample values and the algorithm samples with replacement from the vector or the matrix as the first step of data generation. For two group cases, the user simply specifies \textit{samp.prior.mu.t} and \textit{samp.prior.mu.c} which are vectors of samples for $\mu_t$ and $\mu_c$. For normal data, arguments \textit{samp.prior.var.t} and \textit{samp.prior.var.c}, which contain samples for $\tau_t^{-1}$ and $\tau_c^{-1}$, must also be provided. Example \ref{sec:ex1} demonstrates the use of point mass sampling priors for binary data. 

For GLM cases, the user specifies \textit{samp.prior.beta}, a matrix of samples for $\beta$. For normal data, the argument \textit{samp.prior.var} containing samples for $\tau^{-1}$ must also be provided. For example, suppose one wants to compute the power for the hypotheses $$H_0: \beta_1 \ge 0$$ and $$H_1: \beta_1 < 0.$$ To approximate the sampling prior for $\beta_1$, one can simply sample from a truncated normal distribution with negative mean, so that the mass of the prior falls in the alternative space. Conversely, to compute the type I error rate, one can  sample from a truncated normal distribution with positive mean, so that the mass of the prior falls in the null space. Next, to generate the sampling prior for the other parameters $(\beta_0,\beta_2, \cdots,\beta_p)$, one can use the posterior samples given the historical data as the discrete approximation to the sampling prior. The function \textit{glm.fixed.a0} generates such posterior samples if the  \textit{current} argument is set to \textit{FALSE} and $a_{0k}=1$ for $k=1,\cdots,K.$ Section \ref{sec:ex2} illustrates this method for binary data with covariates. \cite{psioda_2018} discusses sampling prior elicitation in detail.

\subsection{Approximation for GLMs}
Because running \textit{power.glm.fixed.a0} and \textit{power.glm.random.a0} is potentially time-consuming, an approximation method based on asymptotic theory  \citep{ibrahim_2015} has been implemented for the model with fixed $a_0$. In order to attain the exact sample size needed for the desired power, the user can start with the approximation to get a rough estimate of the sample size required, using \textit{power.glm.fixed.a0} with \textit{approximate=TRUE}. Section \ref{sec:ex2} illustrates the use of the approximation method. For normal data, the closed form of the distribution of the MLE of $\beta$ is derived and used to compute power. For other types of data, the Newton-Raphson algorithm is used. Only canonical links are allowed.

\section{Examples} 
\label{sec:ex}
\subsection{Design of a non-inferiority trial for medical devices}
\label{sec:ex1}
We first consider the non-inferiority design application of \cite{Chen_2011} considering a model for binary outcomes for treatment and control groups with no covariates. The goal of that application was to design a trial to evaluate a new generation of drug-eluting stent (DES) (“test device”) with the first generation of DES (“control device”). The primary endpoint is the 12-month Target Lesion Failure (TLF), defined as any of ischemia-driven revascularization of the target lesion (TLR), myocardial infarction (MI) (Q-wave and non-Q-wave) related to the target vessel, or (cardiac) death related to the target vessel. Historical information can be borrowed from two previously conducted trials involving the first generation of DES. The historical data are subsets of the data published in Stone et al. (2004, 2005). Table \ref{summary} summarizes the historical data. 

\begin{table}[t!]
\centering
\setlength{\extrarowheight}{5pt}
\begin{tabular}{cc}
\hline
& 12-Month TLF   \\
& \% TLF (\# of failure/$n_{0k}$)\\
\hline
Historical Trial 1 & 8.2\% (44/535)  \\
Historical Trial 2& 10.9\% (33/304)    \\
\hline
\end{tabular}
\caption{Summary of historical data for the medical devices study.}
\label{summary}
\end{table}

We will illustrate Bayesian SSD incorporating historical data using the power prior with fixed $a_0$ and the normalized power for $a_0$ modeled as random. Let $\textbf{y}_t^{(n_t)}=(y_{t1},\cdots, y_{tn_t})$ and $\textbf{y}_c^{(n_c)}=(y_{c1},\cdots, y_{cn_c})$ denote the responses from the current trial for the test device and the control device, respectively. The total sample size is $n=n_t+n_c$. We assume the $i$-th observation from the test group $y_{ti}$ follows Bern($\mu_t$), and the $i$-th observation from the control group $y_{ci}$ follows Bern($\mu_c$). Note that the notation used in our package is different from the notation used in \cite{Chen_2011}, which assumes $y_{ti}$ follows Bern($p_t$) and $\mu_t=\log\left(\frac{p_t}{1-p_t}\right)$.  The hypotheses for non-inferiority testing are $$H_0: \mu_t - \mu_c \ge \delta$$ and $$H_1: \mu_t - \mu_c < \delta,$$ where $\delta$ is a prespecified non-inferiority margin. We set $\delta=4.1\%$. We choose beta$(10^{-4}, 10^{-4})$ for the initial prior for $\mu_c$, which performs similarly to the uniform improper initial prior for $\log\left(\frac{\mu_c}{1-\mu_c}\right)$ used in \cite{Chen_2011} in terms of operating characteristics. Power is computed under the assumption that $\mu_t=\mu_c$ and type I error rate is computed under the assumption that ${\mu_t=\mu_c+\delta}$. For sampling priors, a point mass prior at $\mu_c = 9.2\%$ is used for $\pi^{(s)}(\mu_c)$ where $9.2\%$ is the pooled proportion for the two historical control datasets, and a point mass prior at $\mu_t = \mu_c$ is used for $\pi^{(s)}(\mu_t)$. For all computations, we use $N=10,000$, $\frac{n_t}{n_c} = 3$, and $\gamma=0.95$. For this example, we consider $n_t=750$ and $a_{01}=a_{02}=0.3$. Power can be calculated with following code in \textbf{BayesPPD}. The \textit{historical} matrix is defined where each row represents a historical dataset, and the three columns represent the sum of responses, sample size and $a_0$, respectively, of the historical control data. Since point mass sampling priors are used for $\mu_t$ and $\mu_c$, \textit{samp.prior.mu.t} and \textit{samp.prior.mu.c} are both scalars. For Bernoulli outcomes, beta initial priors are used for $\mu_t$ and $\mu_c$, with hyperparameters specified by \textit{prior.mu.t.shape1}, \textit{prior.mu.t.shape2}, \textit{prior.mu.c.shape1} and \textit{prior.mu.c.shape2}.

\begin{verbatim}
R> historical <- matrix(0, ncol=3, nrow=2)
R> historical[1,] <- c(44, 535, 0.3)
R> historical[2,] <- c(33, 304, 0.3)
R> 
R> set.seed(1)
R> power <- power.two.grp.fixed.a0(data.type="Bernoulli", 
+    n.t=750, n.c=round(750/3), historical=historical,
+    samp.prior.mu.t=0.092, samp.prior.mu.c=0.092,
+    prior.mu.t.shape1=0.0001, prior.mu.t.shape2=0.0001, 
+    prior.mu.c.shape1=0.0001,prior.mu.c.shape2=0.0001,
+    delta=0.041, N=10000)
R> power$power/type I error
[1] 0.8428
\end{verbatim}

When $a_0$ is random, the normalized power prior is used and the priors for $a_{01}$ and $a_{02}$ are beta(1,1), as in \cite{Chen_2011}. We use the default settings for the upper limits, lower limits and slice widths for $a_{01}$ and $a_{02}$. We run 20,000 iterations of the slice sampler. The same initial priors and sampling priors are used as in the fixed $a_0$ case. The code is shown below for $n_t=750$.

\begin{verbatim}
R> historical <- matrix(0, ncol=2, nrow=2)
R> historical[1,] <- c(44, 535)
R> historical[2,] <- c(33, 304)
R> 
R> set.seed(1)
R> power <- power.two.grp.random.a0(data.type="Bernoulli", 
+   n.t=750, n.c=round(750/3),historical=historical,
+   samp.prior.mu.t=0.092, samp.prior.mu.c=0.092,
+   prior.mu.t.shape1=0.0001, prior.mu.t.shape2=0.0001, 
+   prior.mu.c.shape1=0.0001,prior.mu.c.shape2=0.0001,
+   prior.a0.shape1=1,prior.a0.shape2=1,
+   delta=0.041, gamma=0.95,
+   nMC=20000, nBI=250, N=10000)
R> power$`power/type I error`
[1] 0.864
\end{verbatim}

Table \ref{res} compares power calculations from \cite{Chen_2011} and \textbf{BayesPPD} for a few different sample sizes.

\begin{table}[t!]
\centering
\setlength{\extrarowheight}{5pt}
\scalebox{0.8}{
\begin{tabular}{lcccccc}
\hline
Total sample size&& 1000 & 1080 & 1200 & 1280 & 1480 \\
$n_t$ && 750 & 810 & 900 & 960 & 1110 \\
$n_c$ && 250 & 270 & 300 & 320 & 370\\
\hline
&&Power&&&&\\
$a_0 = (0.3, 0.3)$ &\textbf{BayesPPD} & 0.843 & 0.858 & 0.889 & 0.898 & 0.924 \\
 &\cite{Chen_2011} & 0.840 & 0.856 & 0.884 & 0.892 & 0.923 \\
Random $a_0$ &\textbf{BayesPPD} &  0.864 &  0.885 & 0.909 &  0.921 & 0.937 \\
&\cite{Chen_2011} & 0.843 & 0.878 &  0.897 & 0.902  & 0.914 \\
\hline
&&Type I Error Rate&&&&\\
$a_0 = (0.3, 0.3)$ &\textbf{BayesPPD} & 0.030 & 0.027 & 0.032 & 0.030 &  0.032 \\
 &\cite{Chen_2011} & 0.030  &  0.027 &  0.028 &  0.030 & 0.032 \\
Random $a_0$ &\textbf{BayesPPD} &  0.032 & 0.027  & 0.031 &  0.031 & 0.031  \\
&\cite{Chen_2011} &  0.038 &  0.031 &  0.029 &  0.036 &  0.039\\
\hline
\end{tabular}
}
\caption{Estimated power and type I error rate for non-inferiority design.}
\label{res}
\end{table}

\subsection{Study of acquired immunodeficiency syndrome (AIDS)}
\label{sec:ex2}
Using data from two trials that study the effect of Zidovudine on AIDS, ACTG019 and ACTG036, we will demonstrate how \textbf{BayesPPD} can be used for coefficient estimation as well as power and type I error rate calculation for generalized linear models in designs that incorporate historical data. 

Zidovudine (AZT) is an inhibitor of the replication of the human immunodeficiency virus (HIV). The ACTG019 study was a double-blind placebo-controlled clinical trial comparing AZT with a placebo in adults with asymptomatic HIV who had CD4 cell counts of fewer than 500 per cubic millimeter. The results were published in Volberding et al. (1990). The binary primary endpoint is death or development of AIDS or AIDS-related complex (ARC). For this example we consider four of the measured covariates used, CD4 cell count (x01) (cell count per cubic millimetre of serum), age (x02), treatment (x03) and race (x04). The covariates CD4 cell count and age are continuous, while the others are binary. The ACTG036 study was also a placebo-controlled clinical trial comparing AZT with a placebo in asymptomatic patients with hereditary coagulation disorders and HIV infection. The results were published in Merigen et al (1991). The endpoint and covariates used are the same as those in the ACTG019 trial. Table \ref{actg} summarizes the endpoint and covariates for the two studies. 

%

\begin{table}[t!]
\centering
\setlength{\extrarowheight}{3pt}
\begin{tabular}{lcc}
\hline
& ACTG019  & ACTG036 \\
& (control group) &\\
\hline
No. of patients & 404 & 183 \\
AZT treatment, n (\%) & NA & 89 (48.6) \\
CD4 cell count, mean (SD)  & 332.5 (109.3) & 297.7 (130.5)\\
Age, y; mean (SD) & 34.5 (7.7)& 30.4(11.2)\\
White race, n (\%)& 377 (93.3)& 166 (90.7)\\
Death or ARC, n (\%)& 36 (8.9) & 11 (6.0)\\
\hline
\end{tabular}
\caption{Summary of the ACTG019 trial (control group) and the ACTG036 trial data.}
\label{actg}
\end{table}

First, we standardize age for ease of interpretation and take the log of CD4 cell count count. 

\begin{verbatim}
R> data(actg019)
R> data(actg036)
R> Y0 <- actg019$outcome
R> X0 <- actg019[,-1]
R> X0$age_std <- scale(X0$age)
R> X0$T4_log <- log(X0$T4count)
R> X0 <- as.matrix(X0[,c("age_std","race","T4_log")])
R> 
R> Y <- actg036$outcome
R> X <- actg036[,-1]
R> X$age_std <- scale(X$age)
R> X$T4_log <- log(X$T4count)
R> X <- as.matrix(X[,c("treat","age_std","race","T4_log")])
\end{verbatim}

Suppose we are interested in analyzing the relationship between the outcome and the covariates after incorporating historical information. The code below demonstrates the analysis based on a power prior with $a_0$ fixed at $0.5$ and using only the ACTG019 study data as prior information. 

\begin{verbatim}
R> set.seed(1)
R> historical <- list(list(y0=Y0, x0=X0, a0=0.5))
R> result <- glm.fixed.a0(data.type="Bernoulli",
+                data.link="Logistic", y=Y, x=X,
+                historical=historical, nMC=10000, nBI=250)
R> colMeans(result)
[1]  4.8931870 -0.9459501  0.3645510  0.7201122 -1.4784046
\end{verbatim}

Table \ref{table:coefficients} displays the posterior mean and 95\% credible interval for $\beta$ for four different priors, $a_0$ fixed at $0$, $0.5$, and $1$ and $a_0$ modeled as random with a beta$(1,1)$ prior. There is evidence suggesting a negative association between AZT and death but the evidence is not substantial by common criteria (e.g., posterior probability $>$ 0.95).  

\begin{table}
\begin{center}
\setlength{\extrarowheight}{5pt}
\scalebox{0.7}{
\begin{tabular}{l|rr|rr|rr|rr}
\hline
 & \multicolumn{2}{c|}{$a_0=0$} & \multicolumn{2}{c|}{$a_0=0.5$} & \multicolumn{2}{c|}{$a_0=1$} & \multicolumn{2}{c}{$a_0 \sim$ beta(1,1)}  \\
\hline
Intercept          & $9.14$  & $ [ 3.83; 16.34]$       & $4.89$  & $ [ 1.24;  8.27]$       & $3.95$ & $ [ 0.94;  6.98]$        & $4.39$       &  $ [ 1.41;  7.54]$ \\
AZT                & $-0.15$ & $ [-1.80;  1.42]$          & $-0.95$          &  $ [-2.16;  0.25]$  & $-1.00$ & $ [-2.12;  0.14]$          & $-0.96$          & $ [-2.14;  0.08]$ \\
Age (standardized) & $0.32$ & $ [-0.42;  1.04]$   & $0.36$  & $ [-0.01;  0.74]$           & $0.38 $  & $ [ 0.11;  0.68]$       & $0.38$ & $ [ 0.06;  0.67]$       \\
Race               & $0.36$ & $ [-2.35;  3.23]$            & $0.72$           & $ [-1.10;  2.75]$  & $0.93$  & $ [-0.83;  3.05]$          & $0.73$           &  $ [-0.86;  2.44]$ \\
log(CD4)           & $-2.42$ & $ [-3.61; -1.35]$       & $-1.48$  &$ [-2.04; -0.89]$      & $-1.32$ & $ [-1.78; -0.84]$       & $-1.37$      & $ [-1.91; -0.86]$  \\
\hline
\end{tabular}
}
\caption{Posterior mean and 95\% credible interval for $\beta$ incorporating historical data for the four priors.}
\label{table:coefficients}
\end{center}
\end{table}

For this example we consider designing a new clinical trial that is similar to the historical trial, ACTG019.  We hope to acquire a range of sample sizes that can achieve powers around $0.8$ to test the hypotheses $$H_0: \beta_1 \ge 0$$ and $$H_1: \beta_1 < 0$$ based on the chosen sampling priors. Here, $\beta_1$ represents the treatment effect of AZT. First, we generate the input for \textit{samp.prior.beta}, a matrix of samples for $\beta$ representing a discrete approximation of the sampling prior. For $\beta_1$, we sample from a truncated normal distribution with mean $-0.5$, which is our guess of the effect size of AZT. The distribution is truncated to avoid extreme, implausible values for $\beta_1$. For the other parameters, the sampling prior is fixed at the posterior mean of the parameter given the historical data, which can be easily obtained using \textit{glm.fixed.a0} with \textit{current=FALSE}. We then combine the sampling prior for $\beta_1$ and the other parameters into a matrix, as follows:

\begin{verbatim}
R> library(truncnorm)
R> set.seed(1)
R> historical.sp <- list(list(y0=Y0, x0=X0, a0=1))
R> beta.sp <- glm.fixed.a0(data.type="Bernoulli",
+                         data.link="Logistic", 
+                         historical=historical.sp,
+                         nMC=10000, nBI=250, 
+                         current.data = FALSE)
> nSP <- 10000
> mat.sp <- matrix(rep(colMeans(beta.sp), each=nSP), nrow=nSP)
> beta1.sp <- rtruncnorm(nSP, a=-2, b=-0.1, mean=-0.5)
> samp.prior.beta <- cbind(mat.sp[,1], beta1.sp, mat.sp[,2:4])
\end{verbatim}

Next, we use \textit{power.glm.fixed.a0} with \textit{approximate=TRUE} to obtain a rough estimate of the sample size required to achieve a power of $0.8$. The code below experiments with sample sizes $800$, $1000$ and $1200$. We observe that to reach a power of $0.8$, the sample size should be approximately $800$ when $a_0$ is fixed at $0.5$.

\begin{verbatim}
R> set.seed(1)
R> sample.sizes <- c(800,1000,1200)
R> historical <- list(list(y0=Y0, x0=X0, a0=0.5))
R> results <- NULL
R> for(i in 1:length(sample.sizes)){
+   result <- power.glm.fixed.a0(data.type="Bernoulli",
+                                data.size=sample.sizes[i],
+                                historical=historical,
+                                samp.prior.beta=samp.prior.beta, 
+                                delta=0, gamma=0.95,
+                                approximate=TRUE, N=10000)
+   results <- c(results, result)
+ }
R> results
[1] 0.8037 0.8177 0.8391
\end{verbatim}

Finally, we calculate the exact power using the normalized power prior with $a_0$ modeled as random. The \textit{normalizing.constant} function provides the value for \textit{a0.coefficients} of \textit{power.glm.random.a0}. Since there is only one historical dataset, the \textit{grid} is simply a matrix with one column. The code below demonstrates the usage when sample size is $800$. We run $25,000$ iterations of the slice sampler for each of the $10,000$ simulated datasets. The corresponding power is $0.7936$. Power curves for the four different priors for sample sizes ranging from $750$ to $1200$ are plotted in Figure \ref{fig:power}. The underlying estimated power values are displayed in Table \ref{tab:powers} in the Appendix.

\begin{verbatim}
R> grid <- matrix(seq(0.05,1,by=0.1))
R> historical <- list(list(y0=Y0, x0=X0))
R> a0_coef <- normalizing.constant(grid=grid, historical=historical, 
+                data.type="Bernoulli", data.link="Logistic")
R> result <- power.glm.random.a0(data.type="Bernoulli",
+                        data.link="Logistic", 
+                        data.size=800, historical=historical,
+                        samp.prior.beta=samp.prior.beta,
+                        a0.coefficients = a0_coef, 
+                        delta=0, nMC=25000, nBI=250, N=10000) 
R> result$`power/type I error`
[1]  0.7936
\end{verbatim}

\begin{figure}[h!]
\centering
\includegraphics[scale=0.7]{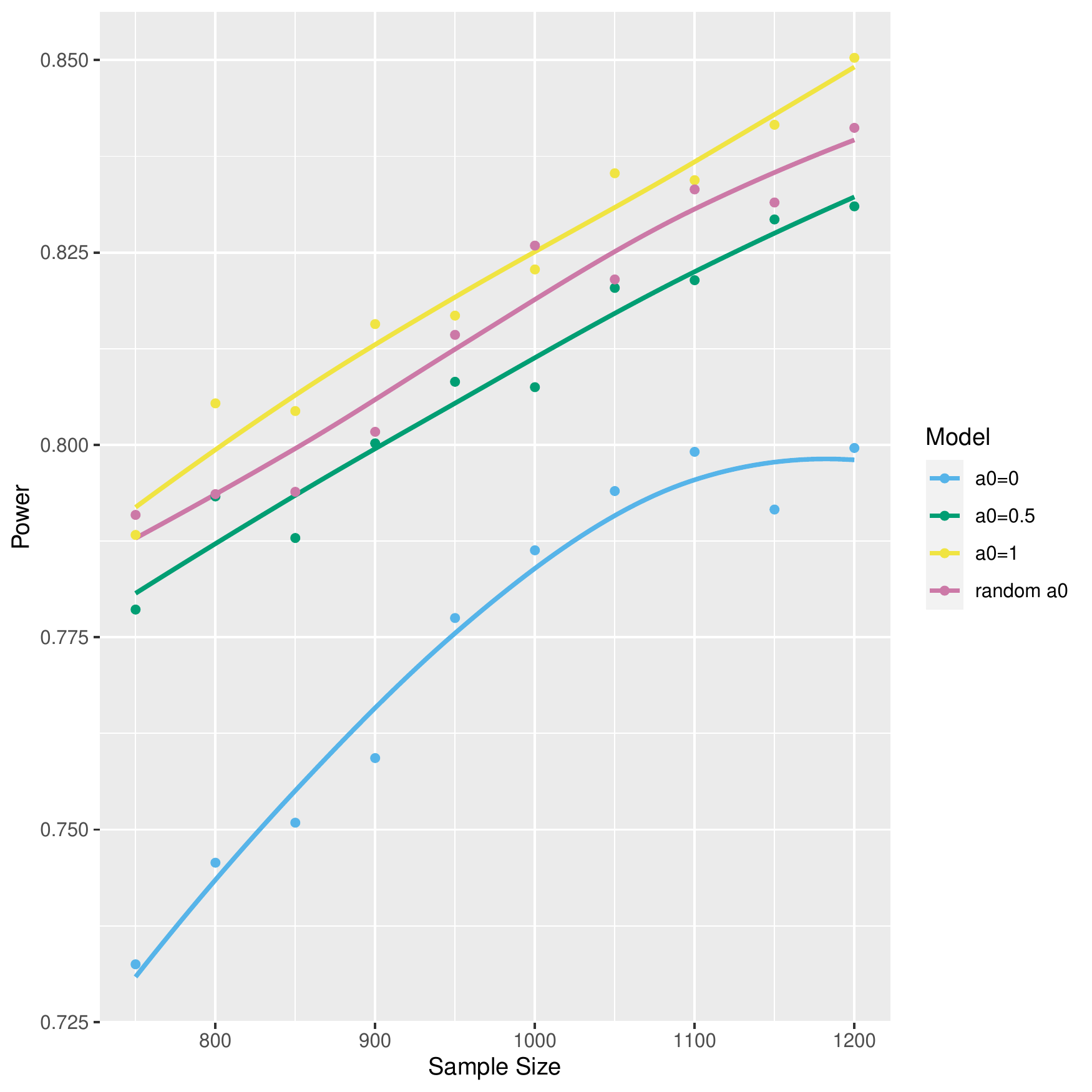}
\label{fig:power}
\caption{Power curves for the four priors. LOESS curves have been fitted to the point estimates. }
\end{figure}

\section{Discussion} 
\label{sec:discussion}

\textbf{BayesPPD} facilitates Bayesian sample size determination by providing a robust suite of functions for power calculation and analysis using the power and normalized power priors for generalized linear models. A major contribution of this package is the ability to handle covariates for Bernoulli, normal, Poisson and exponential outcomes. Despite the use of MCMC algorithms for analysis and design simulations, \textbf{BayesPPD} is computationally efficient, with functions producing results in seconds for many application settings.

A possible extension of the package is the accommodation for longitudinal and time-to-event outcomes. Another potential feature is computing optimal hyperparameters for the beta prior on $a_0$ to ensure certain characteristics are met, such as the ability to adapt to prior-data conflict or prior-data agreement. The method will be based on ongoing theoretical work by the authors.

\newpage

\begin{appendix}

\section{Additional tables} 

\begin{table}[ht]
\centering
\begin{tabular}{ccccc}
  \hline
 Sample size & $a_0=0$ & $a_0=0.5$ & $a_0=1$ & Random $a_0$ \\ 
    \hline
750 & 0.732 & 0.779 & 0.788 & 0.791 \\ 
800 & 0.746 & 0.793 & 0.805 & 0.794 \\ 
850 & 0.751 & 0.788 & 0.804 & 0.794 \\ 
900 & 0.759 & 0.800 & 0.816 & 0.802 \\ 
950 & 0.778 & 0.808 & 0.817 & 0.814 \\ 
1000 & 0.786 & 0.807 & 0.823 & 0.826 \\ 
1050 & 0.794 & 0.820 & 0.835 & 0.822 \\ 
1100 & 0.799 & 0.821 & 0.834 & 0.833 \\ 
1150 & 0.792 & 0.829 & 0.842 & 0.832 \\ 
1200 & 0.800 & 0.831 & 0.850 & 0.841 \\ 
  \hline
\end{tabular}
\label{tab:powers}
\caption{Power for the four priors of the AIDS study.}
\end{table}

\end{appendix}


\end{document}